\documentclass [12pt,twoside]{article}           
\usepackage[left,modulo]{lineno}
\usepackage{setspace}

\countdef\arxiv=201

\ifnum\arxiv=0 \input cpreb.tex \fi

\ifnum\arxiv=1

\usepackage{epsfig,times,lscape}
\usepackage[usenames]{color}

    \ifnum\count102=1

    \fi

    \ifnum\count102=2
\fi

\newcommand{\nc}{\newcommand}

     \ifnum\count101=1
\nc{\qI}[1]{\section{{#1}}}
\nc{\qA}[1]{\subsection{{#1}}}
\nc{\qun}[1]{\subsubsection{{#1}}}
\nc{\qa}[1]{\paragraph{{#1}}}

\def\qpar{\vskip 2mm plus 0.2mm minus 0.2mm}
\def\qL{\hfill \break}
     \fi 

      \ifnum\count101=2
 \nc{\qI}[1]{\parindent=0mm \vskip 8mm 
{\centerline{\LARGE \color{red}#1}}\vskip 3mm}
\nc{\qA}[1]{\vskip 2.5mm \noindent 
{{\bf\large\color{blue}  #1}} \vskip 1mm \parindent=0mm}
 \nc{\qun}[1]{\vskip 1mm \noindent {\sl #1 }\quad }

\def\qL{\hfill \break}
\def\qpar{\vskip 2mm plus 0.2mm minus 0.2mm}

      \fi

\def\qth{\vrule height 12pt depth 0pt width 0pt}
\def\qtb{\vrule height 0pt depth 5pt width 0pt}

\nc{\qfoot}[1]{\footnote{{#1}}}

      \ifnum\count101=1
\def\qbu{\hfill \par \hskip 6mm $ \bullet $ \hskip 2mm}
\def\qee#1{\hfill \par \hskip 6mm (#1) \hskip 2 mm}
      \fi
      \ifnum\count101=2
\def\qbu{\hfill \par \hskip 4mm $ \bullet $ \hskip 2mm}
\def\qee#1{\hfill \par \hskip 4mm (#1) \hskip 2 mm}
      \fi

\def\qparr{ \vskip 1.0mm plus 0.2mm minus 0.2mm \hangindent=10mm
\hangafter=1}

     \ifnum\count101=1 
 
     \fi
     \ifnum\count101=2

  \def\qcitb#1{\noindent \hbox to 102mm{\hfill \small #1} \vskip 1mm}
      \fi

 \def\qpages#1{\count102=0{\loop\advance\count102 by 1
 \null \vfill\eject \ifnum\count102<#1 \repeat}}

\def\qn#1{\eqno \hbox{(#1)}}

\def\qth{\vrule height 12pt depth 0pt width 0pt}
\def\qtb{\vrule height 0pt depth 5pt width 0pt}

\def\qv{\vskip 0.1mm plus 0.05mm minus 0.05mm}
\def\qhu{\hskip 0.6mm}
\def\qhv{\hskip 3mm}

\def\qhw{\hskip 1.5mm}
\def\qleg#1#2#3{\noindent {\bf \small #1\qhw}{\small #2\qhw}{\it \small #3}\qv }
\fi

\begin{document}
\thispagestyle{empty}



\markboth{{\sl \hfill  \hfill \protect\phantom{3}}}
        {{\protect\phantom{3}\sl \hfill  \hfill}}

\color{yellow} 
\hrule height 20mm depth 10mm width 170mm 
\color{black}
\vskip -17mm

 \centerline{\bf \Large Predictive implications of Gompertz's law.}
\vskip 2mm
 \centerline{\bf \Large }
\vskip 10mm
\centerline{\large 
Peter Richmond$ ^1 $ and Bertrand M. Roehner$ ^2 $
}

\vskip 8mm
\large

{\bf Abstract}\quad 
Gompertz's law tells us that for humans above the age of 35
the death rate increases exponentially with a doubling time
of about 10 years.
Here, we show that the same law continues 
to hold even for ages over 100. 
Beyond 106 there is so far no statistical
evidence available because 
the number of survivors is too small even in the largest nations.
However assuming that Gompertz's law continues
to hold beyond 106, we conclude that
the mortality rate becomes equal to 1 at age 120
(meaning that there are 1,000 deaths in a population
of one thousand).
In other words, the upper bound of human life is near 120.
The existence of this fixed-point has interesting implications.
It allows us to predict the form of the relationship
between death rates at age 35 and the
doubling time of Gompertz's law. In order to test this prediction,
we first carry out a transversal analysis for a sample
of countries comprising both industrialized and developing nations.
As further confirmation, we also develop a longitudinal analysis using
historical data 
over a  time period of almost two centuries.
Another prediction arising from this fixed-point model,
is that, above a given population
threshold, the lifespan of the oldest person is independent
of the size of her national community.
This prediction is supported by available empirical evidence.

\vskip 5mm
\centerline{\it Version of 25 September 2015. }

\vskip 1mm
{\small Key-words: death rate, Gompertz's law, oldest persons,
transversal analysis, longitudinal analysis.}
\vskip 5mm

{\normalsize 
1: School of Physics, Trinity College Dublin, Ireland.
Email: peter\_richmond@ymail.com \qL
2: Institute for Theoretical and High Energy Physics (LPTHE),
University Pierre and Marie Curie, Paris, France. 
Email: roehner@lpthe.jussieu.fr
}

\vfill\eject

\qI{Introduction}

Among social phenomena there are very few that are governed by
laws which are valid with good accuracy
in all times and all countries. Gompertz's law is one of them.
For a physicist Gompertz's law is fairly unusual because
it is an exponential change whose rate itself changes
in an exponential way. One will not be surprised that 
such a process reaches a critical point within a finite time.
The present paper draws several implications from this observation%
\qfoot{The present paper is the third in a comparative
biodemographic investigation which, so far, 
comprised the following steps: Richmond and Roehner (2015 a,b).}%
.
\qpar

In 1825
Benjamin Gompertz (1779--1865) derived the law named
after him from life tables 
for the cities of Carlisle and Northampton. Gompertz's law states
that for ages over $ t_1=35 $, the mortality
rate $ \mu(t)=(1/s)ds/dt $, where $ s(t) $ denotes the
population of a cohort in the course of time,
increases in an exponential manner: 
$$ \mu(t)=d_1 \exp\left[\alpha (t-t_1)\right] 
\quad d_1:\hbox{ death rate at age } t_1\simeq 35\hbox{ years} $$

If this law also holds in old age,
an obvious implication is that the extinction of a population
does not occur asymptotically, but within a finite time interval.
This is a consequence of the fact that 
if for an age $ t_2 $ the death rate
$ \mu(t_2)=(1/s)\left[\Delta s/\Delta t\right] $ in a unit time
interval $ \Delta t =1 $ reaches 1, then the number of deaths 
$ \Delta s $ equals the population $ s(t) $ which means that the
population vanishes for $ t=t_2 $. In other words,
$ t_2 $ represents the strict upper bound of the population's life time. 
Just to show that this is not purely theoretical, we note
that for the population considered in the study by Arthur Roger
Thatcher and his collaborators (1999), 79\% of the males aged 109
died before reaching 110.
\qpar

The data for Gompertz's law up to age 106 are summarized
in Fig. 1a. The data come from two sources. For ages up to 80 
(or sometimes 95) 
the death rates can be taken from standard death rate
statistics as published in all countries. For ages up to 106,
the data are taken from Thatcher (1999) based on a
global sample of 13 countries over several years.
The monograph actually
gives death rates up to age 113, but beyond age 106 the numbers
involved become very small
which gives rise to substantial random fluctuations.
\qpar
Incidentally, such fluctuations
explain why studies using national death rate data for persons over
100 lead to fairly shaky results.
Some studies (e.g.
Strehler 1967) display a deceleration effect,
whereas in others (e.g. the graph for the United States 
in 2003 that accompanies
the Wikipedia article entitled ``Gompertz-Makeham law of mortality'')
there appears to be a death rate acceleration.
For the age interval 80-106 Thatcher et al. (1999) show clearly that
Gompertz's law
continues to hold.
Beyond 106 there is still no convincing evidence. One will have to wait
until studies similar to the Thatcher study are performed in
countries with large populations such as China or India.
\qpar

\begin{figure}[htb]
\centerline{\psfig{width=16cm,figure=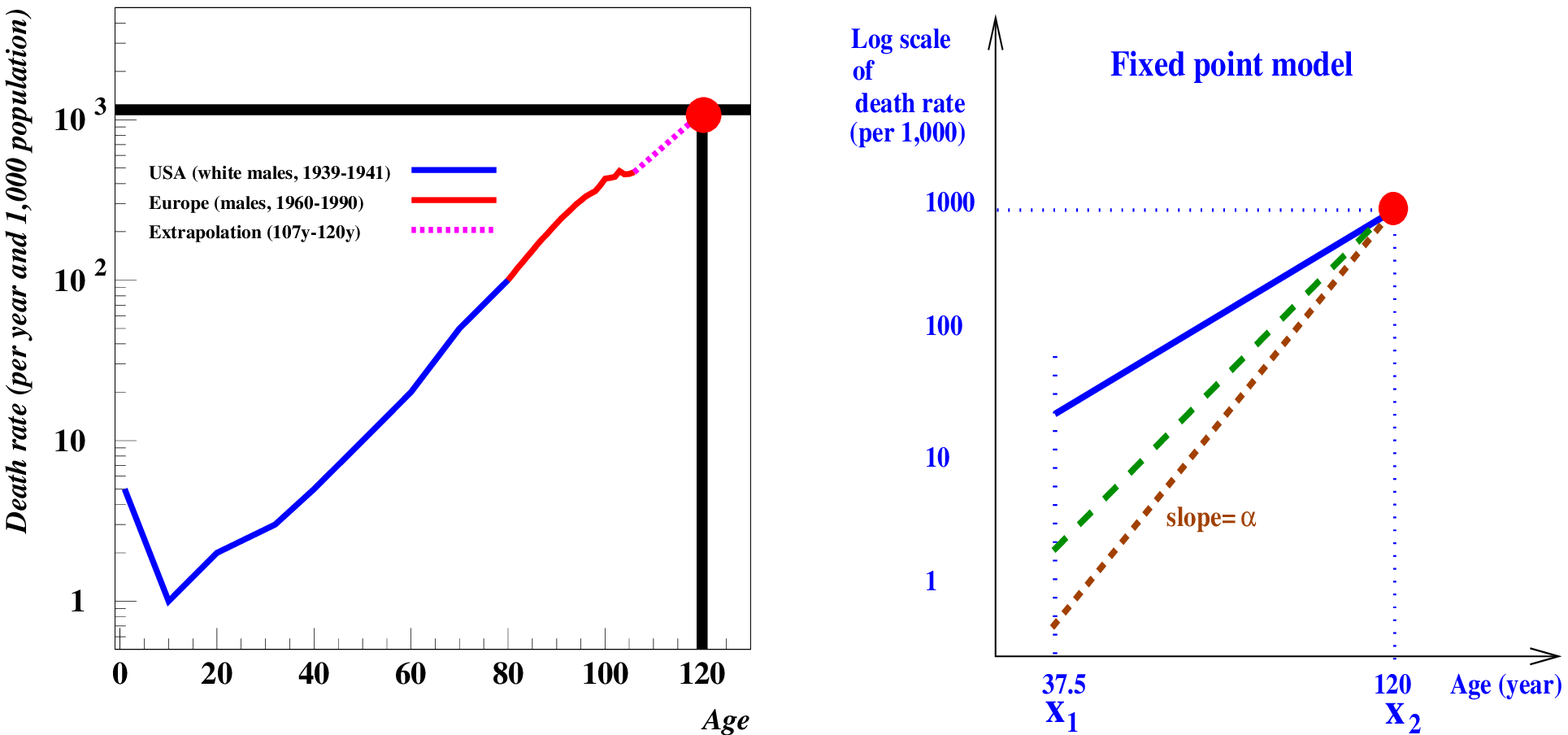}}
\qleg{Fig. \qhu 1a,b\qhv Gompertz's law.}
{{\bf Left}: Gompertz's law is characterized by a death rate which
increases exponentially with age: $ \mu(t)=d_1 \exp(\alpha (t-t_1)) $.
Here $ t_1 $ is about 35 and for $ t>t_1 $ one gets
$ \alpha=0.074 $ which corresponds to a doubling time
of $ \theta=\log2/\alpha=9.4 $ years. The empirical part of the
death rate curve relies on two different sources: 
(i) first, standard vital statistics; 
usually such data do not go beyond 95. 
(ii) the data collected by Arthur Thatcher and
his collaborators had for objective to cover ages beyond 95.
Yet, beyond the age of 107, the samples become too small
even for very large initial populations.
That is why,
for ages between 107 and 120, we make the conjecture that the
line can be extrapolated. This extrapolation leads to the fact that
approximately at age 120 the
death rate reaches the value 1,000 per year and per 1,000
persons which means that the population vanishes. This upper bound
of 120 agrees fairly well with verified (worldwide)
maximum lifespans: Jeanne Calment (122) and Sarah Knauss (119).
{\bf Right}: This graph is a schematic illustration
of the fixed-point model. The three Gompertz lines are
supposed to correspond to
different countries and time periods. Fig. 5a displays
a graph of that kind that is based on real data.}
{Sources: Strehler (1967), Thatcher et al. (1999)}
\end{figure}

Fig. 1b illustrates the simple idea upon which the present study is based.
We refer to it as the ``fixed-point model''. 

If over age $ t=35 $ Gompertz's law holds for all countries and
if 120 years is the end point of human life, then the equation
of any straight line which summarizes Gompertz's law for
a specific country
will be determined by the initial death rate from which it starts.
If one denotes the logarithm of the death rate  by $ y $,
according to a standard formula, 
this equation reads:
 $$ y=\alpha (t-t_1) +q,\quad
\alpha= { y_2-y_1 \over t_2-t_1 },\
q = { t_2y_1-t_1y_2 \over t_2-t_1 }\quad 
y=\log(\hbox{\normalsize death rate}),\ t=\hbox{\normalsize age} $$

For the initial age $ t_1 $ we selected an age such that in
any country
Gompertz's law holds for $ t \geq t_1 $. Observation shows
(see Fig. 2a,b)
that this is the case for the age intervals above (and including)
$ 35-39 $. This leads to a value of $ t_1 $ equal to the
mid-point of this interval, i.e. $ t_1=37.5 $ years.
Then , replacing $ t_1, t_2 $ by 37.5 and 120, and $ y_2 $ by $ \log 1000 $,
one gets the following relationship between the exponent $ \alpha $
and $ \log d_1 $.
$$ \alpha={ 1 \over t_2-t_1 }\log d_1 + { \log 1000 \over t_2-t_1 }=
-0.0121\log d_1 + 0.084 \qn{1} $$ 

In the numerical form of this relationship it is supposed
that $ \alpha $ is 
expressed in $ \hbox{year}^{-1} $ and $ d_1 $ in number of 
deaths per year and per 1,000 population.

\qI{Gompertz's exponent's changes and the filter effect}

\qA{Transversal analysis}
In order to study the variability of Gompertz's law 
across countries, we need
to set up an appropriate ``experiment''. In other words
we need to define the kind of data that we need.

\qun{What data do we need?}

Firstly, because we need data for several countries
we must use a source providing such international data.
The Demographic Yearbooks published by the United
Nations serve our needs well.
\qpar
Secondly, in order to be able to deduce
the relationship between $ d_1 $ and $ \alpha $ we need observations
for a wide range of $ d_1 $.
We cannot restrict ourselves to
data from industrialized countries because in all these
countries $ d_1 $ will be limited to a narrow
range of small values, basically of the order of
1 per 1,000. 
\qpar
Moreover, since we wish to study the convergence of death
rates toward the fixed-point at age 120, we would
like death rates for age groups which are as close as
possible to 120. Since at this point we are not particularly interested
in the difference between male and females we shall use data for both sexes.
\qpar

Now that we have defined what we need, let us see
what the UN Yearbooks can offer.
By accessing the Yearbook of 2011
one observes immediately two things,
\qbu While {\it numbers}
of deaths by age are given for many countries,
death {\it rates} are given for only about one third
of the countries.
\qbu In the subset of countries for which rates
are available only about one third of them provide data
up to the age group 95-99. For the others, the
data are given only up to 70, 80 or 85 depending
on the country. This will lead us to distinguish
two groups of countries: group $ A $ which contains
all countries providing data up to age-group
95-99 (such data will be referred to as ``complete data'')
and group $ B $ which will contain all other countries
(it will be referred to as the ``incomplete data'' group). 
\qbu In group $ A $ there are countries with
death rates at age 37.5 that are as low as 0.6 per 1,000.
At the other end of the spectrum
one would expect to find high death rates in developing countries
and particularly in African countries. However, among the
African countries with sizable populations there are only 4 for
which death rates are given, namely: Egypt, Sierra Leone,
Swaziland and Zimbabwe. Egypt was included in group $ B $ but
its death rate at age 37.5 is $ d_1=1.7 $ which is not particularly
high. In the other three countries $ d_1=11.3,\ 36.3,\ 38.9 $.
Such values would be quite useful but unfortunately
the death rate series do not appear reliable. This can
be seen in two ways. First, by the fact that they are
labeled by the UN 
as I (instead of C which means ``complete'') and secondly
because the death rates do not increase in a monotonic way. 
Thus, for Swaziland the death rates in the age groups
$ 70-74 $ and $ 75-79 $ are $ 48.0 $ and $ 45.4 $ respectively
which, although not altogether impossible, is highly unlikely.
\qpar

In conclusion, we expect reliable results in group $ A $ but
fairly poor results in group $ B $.

\qun{Results}

For each and every country, Gompertz's law holds with high
accuracy (see Fig. 2a,b). More precisely, 
when reliable statistics are used
the correlation 
(age, logarithm of death rate) is always higher than 0.995.
Yet, the slopes may differ substantially. This leads to the two
following  questions.
\qee{1} Is there a regularity in the variations of the slopes
or is it just random?
\qee{2} If there is a regularity does it follow the
fixed-point model?
\qee{3} If the data are indeed well described by the
fixed-point model, how can one explain this effect
in biological terms?
\qpar

Fig. 3a,b answers the questions 1 and 2.

\begin{figure}[htb]
\centerline{\psfig{width=14cm,figure=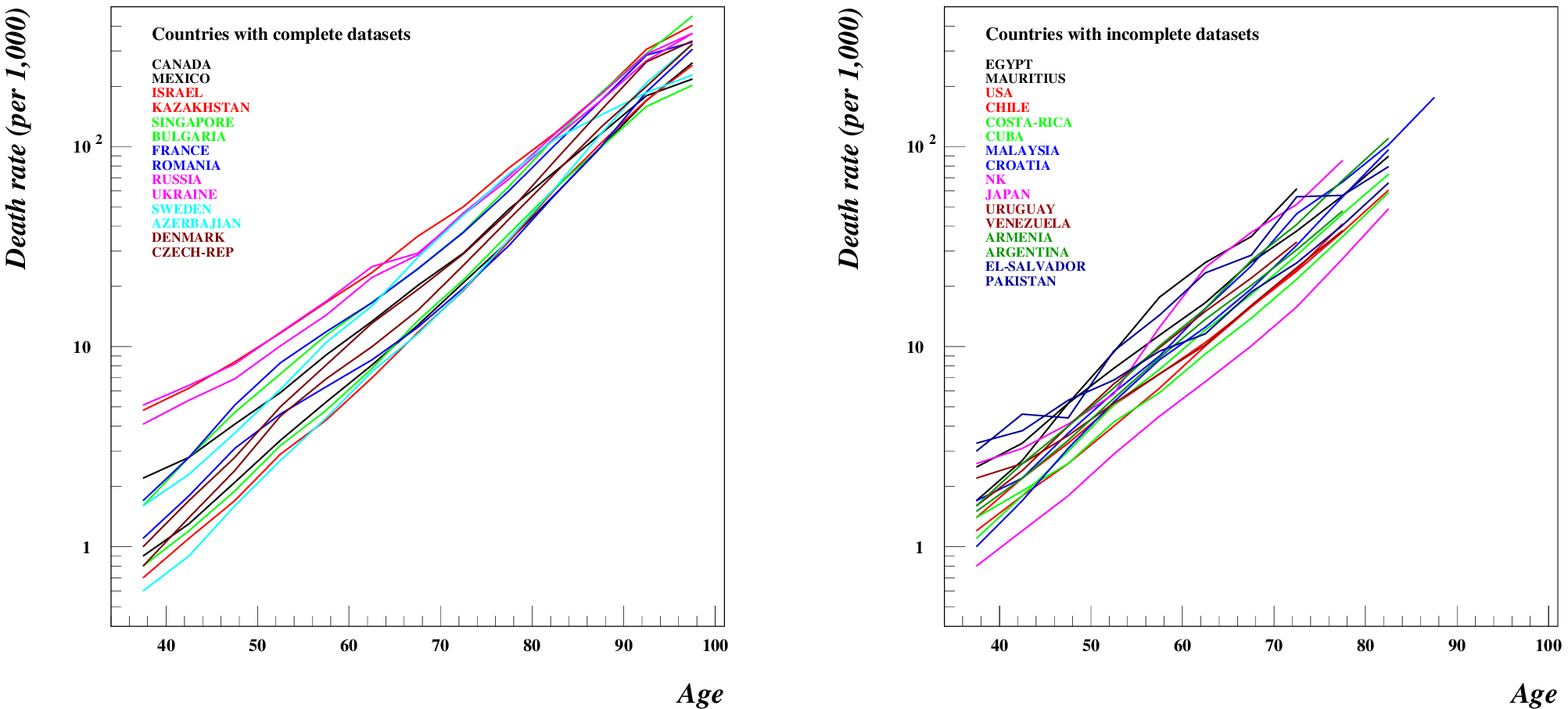}}
\qleg{Fig. \qhu 2a,b\qhv Gompertz's law in two sets of countries.}
{{\bf Left}: Group ($ A $) of
countries which give death rates for ages
up to 95-99 year old. {\bf Right}: Group ($ B $ ) of
countries which provide death rate
data only for shorter age intervals. In Fig. 3a,b and 4a,b 
complete and incomplete data will also be on the left and
right respectively.
Ideally all lines are expected
to converge toward the end-point (120,1000). Not surprisingly,
this convergence is clearer in group $ A $ than in group $ B $.}
{Source: United Nations Demographic Yearbook, 2011.}
\centerline{\psfig{width=14cm,figure=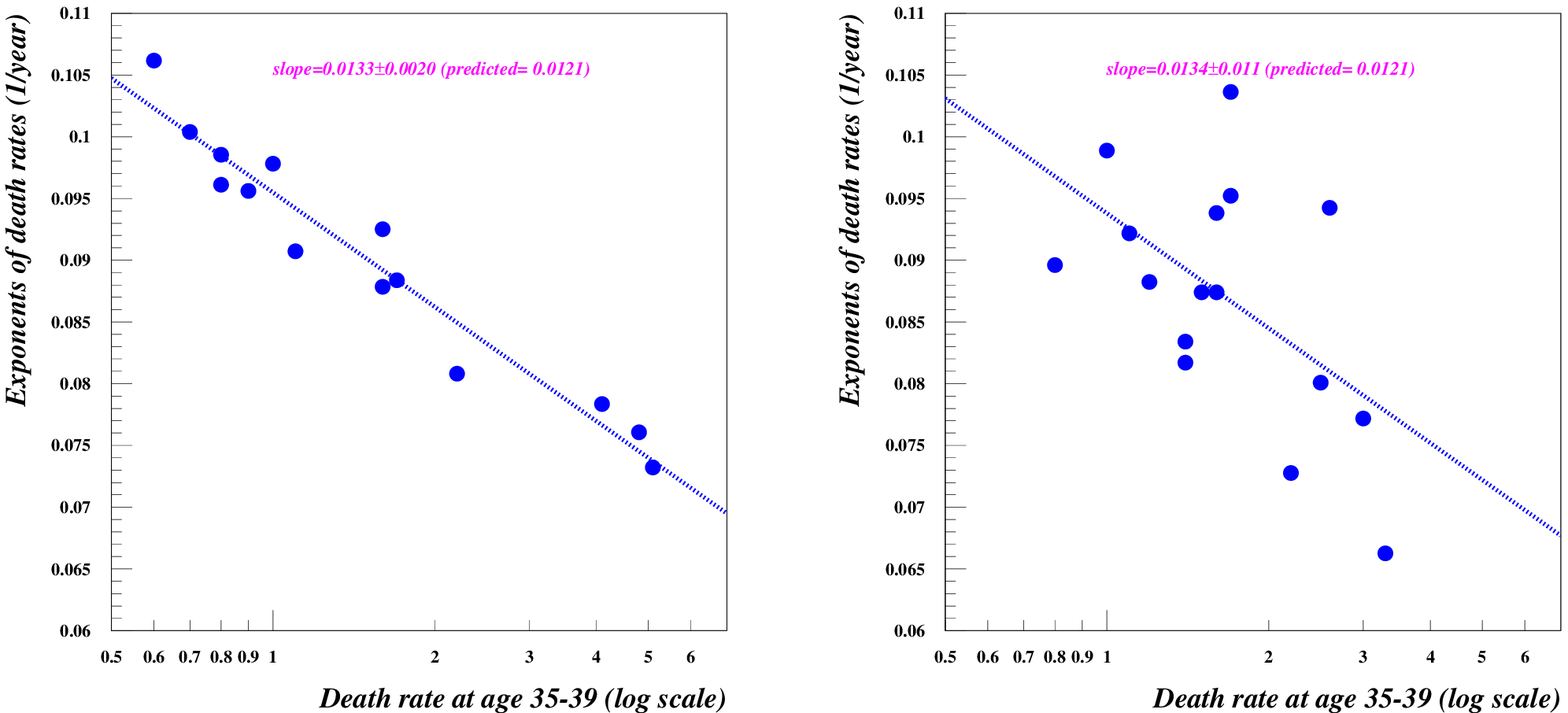}}
\qleg{Fig. \qhu 3a,b\qhv Exponents of Gompertz's law as a function
of death rates in the $ 35-39 $ age-group ($ \log d_1 $).}
{{\bf Left}: Group $ A $ of countries with complete datasets.
The correlation $ (\log d_1,\alpha ) $ is $ -0.970 $. 
{\bf Right}: Group $ B $ of countries with incomplete datasets.
The correlation is $ -0.55 $. 
Although the slopes of the regression lines are almost the same
in the two groups,
the accuracy of their measurement 
is about 5 times better in group $ A $ than in group $ B $.
The exponents predicted by the fixed-point model
are within the error bars of the observations.}
{Source: United Nations demographic yearbook, 2011.}
\end{figure}
%
\begin{figure}[htb]
\centerline{\psfig{width=12cm,figure=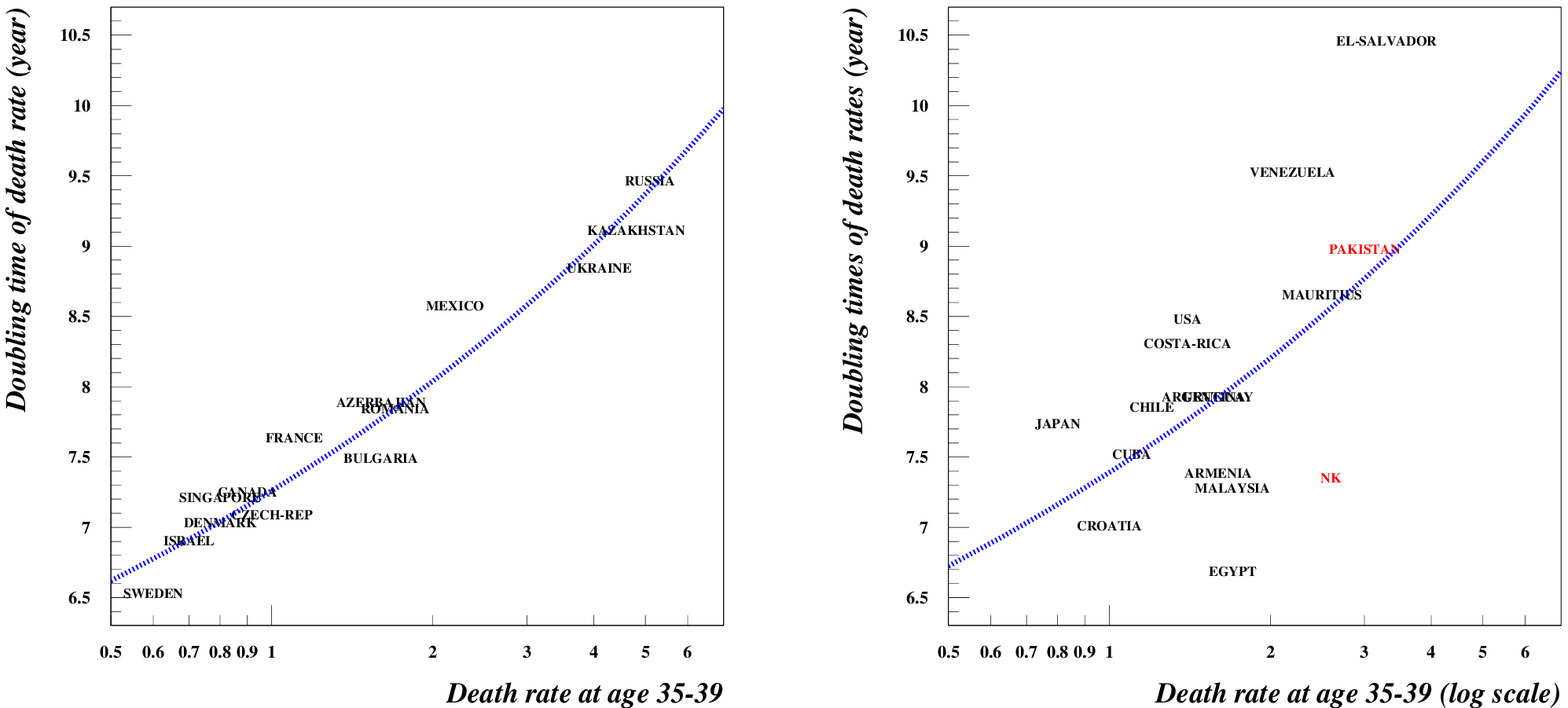}}
\qleg{Fig. \qhu 4a,b\qhv  Doubling times $ \theta $ of death rates
as a function of death rates in the initial
$ 35-39 $ age-group ($ d_1 $).}
{{\bf Left}: Group $ A $ of countries with complete datasets.
{\bf Right}: Group $ B $ of countries with incomplete datasets.
According to the fixed-point model there should be an
hyperbolic relationship between the two variables:
$ \theta=\log 2/(-a\log d_1+b) $. The non linearity
of the relationship would become more obvious for 
higher initial death rates. In some African countries
(e.g. Sierra Leone or Zimbabwe) initial death rates
as high as 30 per 1,000  were reported
but the demographic data of such countries
appear fairly unreliable (see text).
The names of North Korea and
Pakistan are written in red because the data provided
by these countries are acknowledged to be incomplete
(even for the restricted age range for which data are
available).}
{Source: United Nations Demographic Yearbook, 2011.}
\end{figure}

More precisely one
gets the results summarized in Table 1.

\begin{table}[htb]

\small

\centerline{\bf Table 1\quad Relationship between initial
death rates $ d_1 $ and Gompertz's exponents: 
$ \alpha =-a\log d_1+b $}

\vskip 5mm
\hrule
\vskip 0.7mm
\hrule
\vskip 2mm

$$ \matrix{
\qtb
\hbox{}\hfill & 100\times a & 100\times b \cr
\noalign{\hrule}
\qth
\hbox{Prediction of the fixed-point model }\hfill &  1.21&  8.37 \cr
\hbox{}\hfill &  & \cr
\hbox{Group A (complete data)}\hfill & 1.33\pm 0.20 &
9.54\pm 0.13 \cr
\hbox{Difference with respect to prediction}\hfill & 10\% & 14\%\cr
\hbox{}\hfill &  & \cr
\hbox{Group B (incomplete data)}\hfill & 1.34\pm 1.0
& 9.37\pm 0.41 \cr
\qtb
\hbox{Difference with respect to prediction}\hfill & 11\% & 12\%\cr
\noalign{\hrule}
} $$
\vskip 1.5mm
\small
Notes: For clarity the table gives $ a, b $ multiplied by one hundred.
Comparison of the error bars (probability level of 95\%)
shows that for $ a $ as well as $ b $ the
accuracy of the observations in group $ A $  is five times
better than in group $ B $. 
{\it }
\vskip 5mm
\hrule
\vskip 0.7mm
\hrule
\end{table}

\qA{Longitudinal analysis}

Western countries began to
collect reliable demographic statistics in the mid-19th century.
As death rates at age 35-39 were substantial higher at that time
can we possibly use such data to explore the region
of high $ d_1 $ that was out of reach in our cross-national
analysis? As a test-case we consider France.
In INSEE (1966) one can find death rate data by age-group
from 1806 on. 
$$ 1806:\ d_1=14.8,\quad 1836:\ d_1=11.8,\quad 1866:\ d_1=11.3,\quad 
2005:\ d_1=1.1 $$

As the data for 1806 may be somewhat less reliable than
later ones, we selected: 1836, 1866 and 2005. Our purpose in taking
1836 and 1866 in spite of the fact that they have almost
the same $ d_1 $ is to control the reliability of the data.
\qpar

This longitudinal analysis leads to the results
summarized in Fig. 5a,b.

\begin{figure}[htb]
\centerline{\psfig{width=14cm,figure=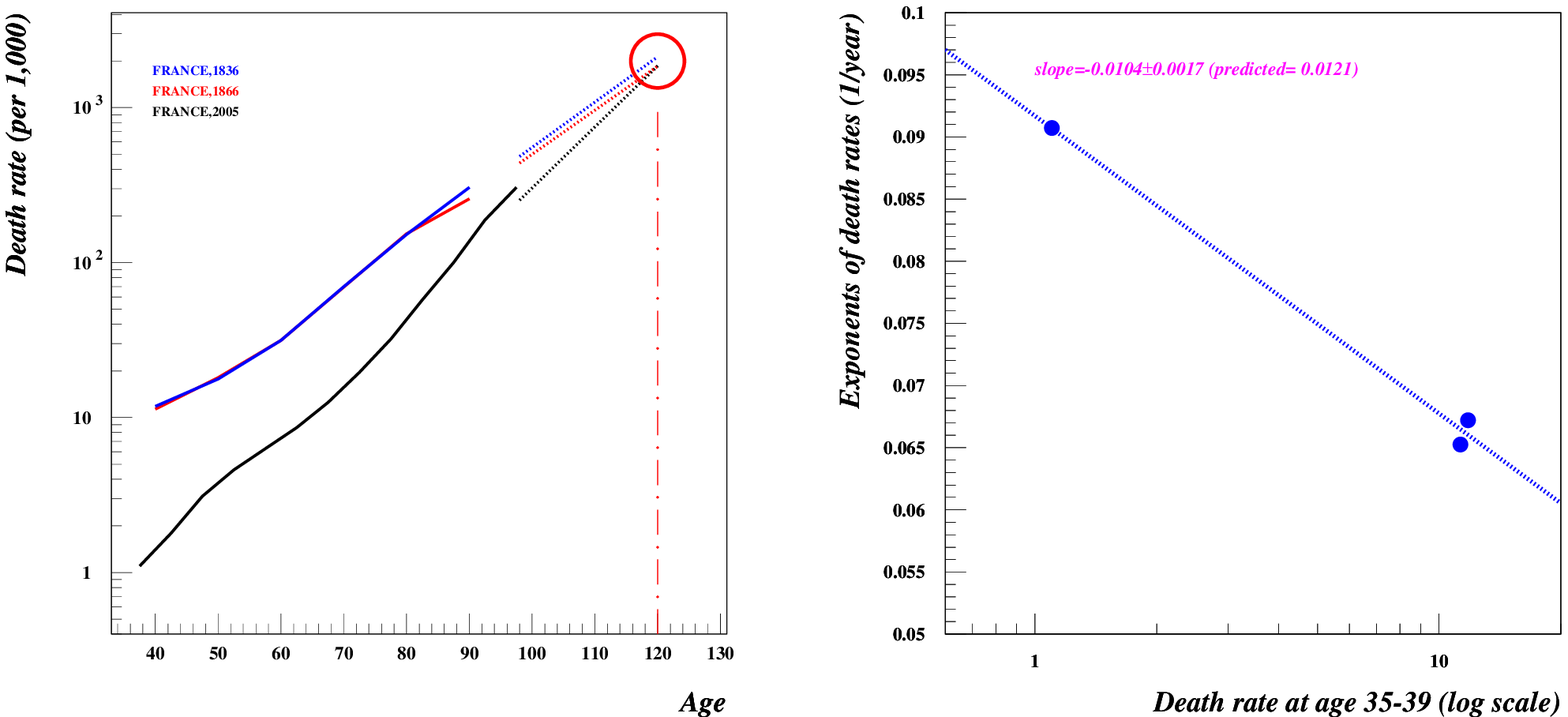}}
\qleg{Fig. \qhu 5a,b\qhv Gompertz's law in France in different
years.}
{{\bf Left}:  The dotted straight lines beyond 95
are the regression lines.
Ideally one expects all lines
to converge toward the end-point (120,1000).
In fact, the center of the red circle around the convergence point
is somewhat higher than 1,000. The fact that the lines
for 1836 and 1866 are almost identical controls the reliability
of the data.
{\bf Right}: Regression line for $ (d_1,\alpha) $ and
comparison with the predicted slope.}
{Source: 2005: United Nations demographic yearbook, 2011;
1836, 1866: INSEE (1966)}
\end{figure}

\qA{Interpretation}

The simplest interpretation which comes to mind for
the results described in the previous subsection 
relies on a filter effect. There are two steps in this explanation.
First, we note that the death rate at age 37.5 is 
not an isolated number but is related to the rates
in other age intervals. Thus, for a sample of countries
the infant mortality rate $ d(0-1) $ and $ d(35-39) $
have a cross-correlation of the order of $ 0.7 $. Similarly,
$ d(35-39) $ has a correlation of same magnitude with $ d(1-4) $.
This means that 
a high $ d(35-39) $ was preceded by a range of fairly high
death rates at a younger age.
It should be observed that intuitively
such correlations are rather unexpected since for 
for such age intervals
the causes of death are fairly different. For the young, say for age below
5, the main causes of death are diseases whereas for the age group 35-39
the causes of death are mostly external factors,  
e.g. accident, suicide, homicide. The fact that there is 
nevertheless a correlation is probably related to
the nature of the social organization. Low infant mortality 
implies a rich country which in turn implies well organized transports
(few traffic accidents) and little violence.
\qpar

Now, the meaning of the filter effect becomes clear.
In a society characterized by a high $ d(35-39) $ there is also
a high infant mortality which means that persons who for diverse
reasons are ``fragile'' (e.g. low immunity, heart malformation)
will die in the early years of their lives.
It follows that the survivors will be less susceptible  
to the diseases which usually come with age which means a
death rate which will increase fairly slowly with age.

\qI{Wall effect for the high end of the lifespan distribution}

The wall effect was already illustrated in Fig. 1 by the
fact that the survivorship function becomes strictly equal
to zero for age $ t=120 $. 
Here, however, the implication will be stated in the 
language of probability theory. The survivorship curve
describes the population of a cohort in the course of time.
If the initial population is normalized to 1, this curve becomes
equivalent to the probability for an individual to reach age $ t $.
In other words it is: $ P\{ X\geq t\}=1-F(t)=G(t) $, where $ F(t) $ is the
cumulative probability distribution function and $ G(t) $ the
complementary distribution function.
Thus,
$ f(t)=-dG/dt $ where $ f(t)dt=P\{t\le X\le t+dt\} $ is the
probability density function. It gives the number of people in
a cohort whose lifespan is in the interval $ (t,t+dt) $.
The function $ G(t) $ for the lifespan in France is given in Fig. 6a.
\qpar
\begin{figure}[htb]
\centerline{\psfig{width=14cm,figure=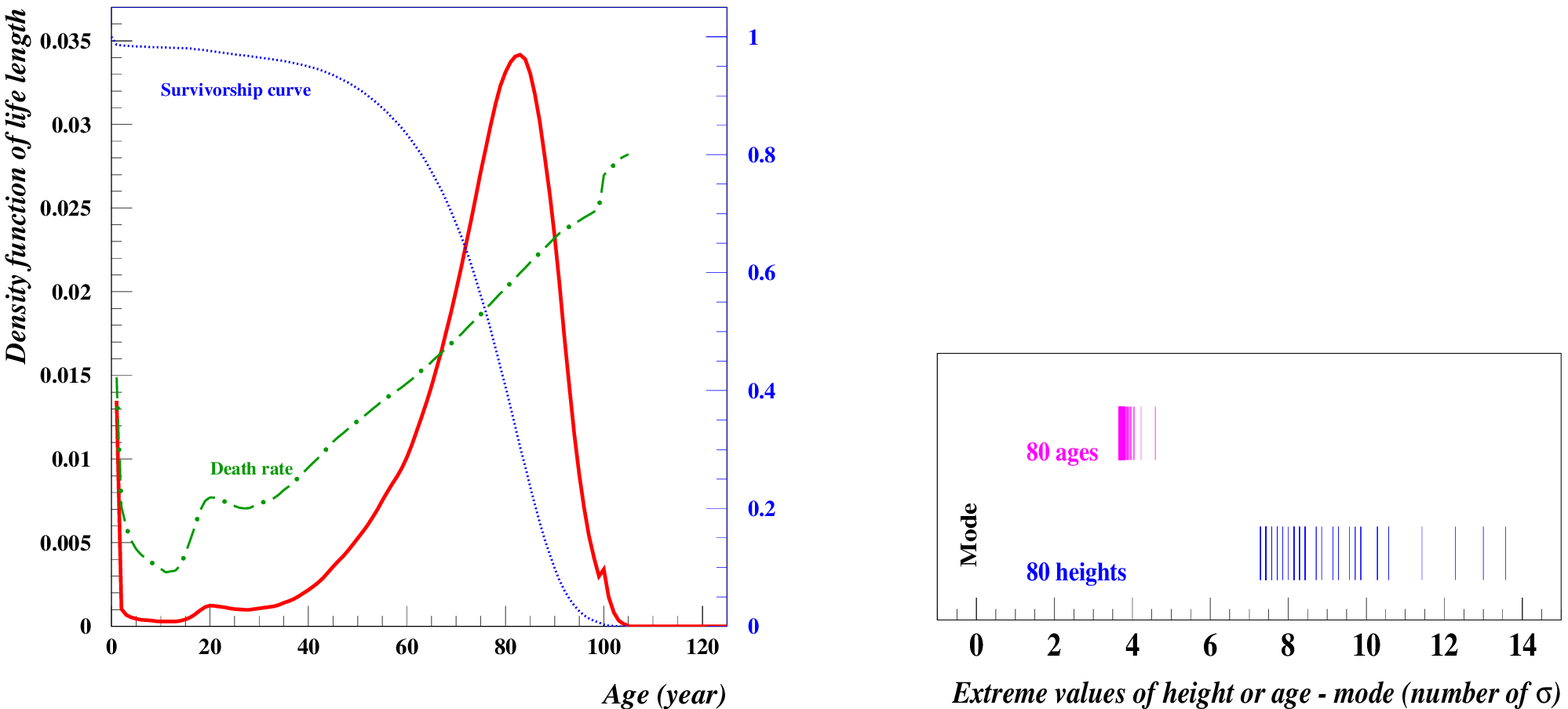}}
\qleg{Fig. \qhu 6a,b\qhv Comparison of
extremes values of height and age.}
{{\bf Left}: Density function of the duration of human life
(both sexes together). 
The survival curve $ s(t) $ is identical to
the complementary distribution function:
$ s(t)=1-F(t) $.
Thus, the density function is the 
opposite of the derivative
of the survivorship curve: $ f(t)=dF/dt=-ds/dt $.
The death rate curve
is the opposite of the derivative
of the logarithm of the survivorship curve: 
$ \mu(t)=(1/s)ds/dt==-d\log[s(t)]/dt $.
{\bf Right}: Comparison between 80 top values for height 
(men) and length of life (women). The vertical lines
show the differences (expressed in number of $ \sigma $)
between individual top values and the mode 
(i.e. peak value) of the distribution.
In each case the standard deviation $ \sigma $ was
computed from the values in the vicinity of the peak.
The wall effect for maximum ages is quite apparent.
Note that for height the small number of vertical lines is due to
the fact that
many realizations correspond to identical positions
because heights are expressed in centimeters rather than in millimeters.
Incidentally, it can be noted that the extreme
values of height do {\it not} follow a Gaussian distribution
for in a real Gaussian (with here $ \sigma\simeq 7 $ cm)
existence of heights of 230cm (i.e. $ 7.6 \sigma $)
would require a population of $ 1/G(7.6)=10^{14} $ people.
The same observation can be made regarding the existence of dwarfs.
}
{Source: Survivorship curve: Official mortality table (TD73-77)
used by insurance companies in France. Extreme values:
Wikipedia lists of tallest and oldest persons.}
\end{figure}

The fact  that: $ f(t)\equiv 0 $ for any $ t\geq 120 $
has two consequences.
\qbu {\color{blue} Maximum values of lifespan} will be
squeezed within a short interval
ending at 120. For the purpose of comparison
Fig. 6b shows also maximum values for height. We chose
this variable because it is usually considered as the archetype of
a Gaussian distribution. 
Incidentally, while indeed true for
heights within 2 or 3 $ \sigma $  around the mean this is no longer
true for the tails of the distribution.
\qbu {\color{blue} Extreme lifespan values are independent
of sample size.}\quad
For any random variable whose distribution function
$ G(t) $ tends toward 0 (yet without reaching 0) as $ t $ increases,
the largest realizations are conditioned by the size of the sample.
Thus, if for the height (expressed in cm) one has
$ G(200)=10^{-3} $ one will need a sample of at least
1,000 individuals in order to have a chance to get one person
with a height over 200 cm. With this logic (and
leaving aside genetic factors) the height of the
tallest persons will be larger in the United States
than in Iceland whose population is 1,000 times smaller.
On the contrary, for age, above an appropriate threshold $ p_0 $,
the size of the sample will not play any role.
This prediction of the fixed-point model is easy to check.
Tables of oldest persons by country can be found on 
Wikipedia. 
Here are some cases. The first number gives
the population $ P $ in millions, 
the second the age $ t $ of the oldest person.
\qpar
Belgium: 11, 112.5; Denmark: 5.6, 115.7; France: 66, 122.4
Germany:  80, 115.2;
Iceland: 0.33, 109.8; Ireland: 4.6, 113.4; Italy: 60, 115.7; 
Moldova: 3.5, 114.8; Russia: 143, 113.0; United States: 319, 119.3.
\qpar
There is a correlation of about 0.70 between $ \log P $ and $ t $ 
but it vanishes for $ t> 10 $ million which means that 
$ p_0\sim 10 $ million.

\qI{Conclusion}

\qA{Extension of the fixed point model to other species}

The fixed-point model 
developed in this
article is based on two observations.
\qee{1} The validity of Gompertz's law for ages over 40.
\qee{2} The fact that the 
upper bound of human life is about 120 years.
\qpar

\begin{figure}[htb]
\centerline{\psfig{width=8cm,figure=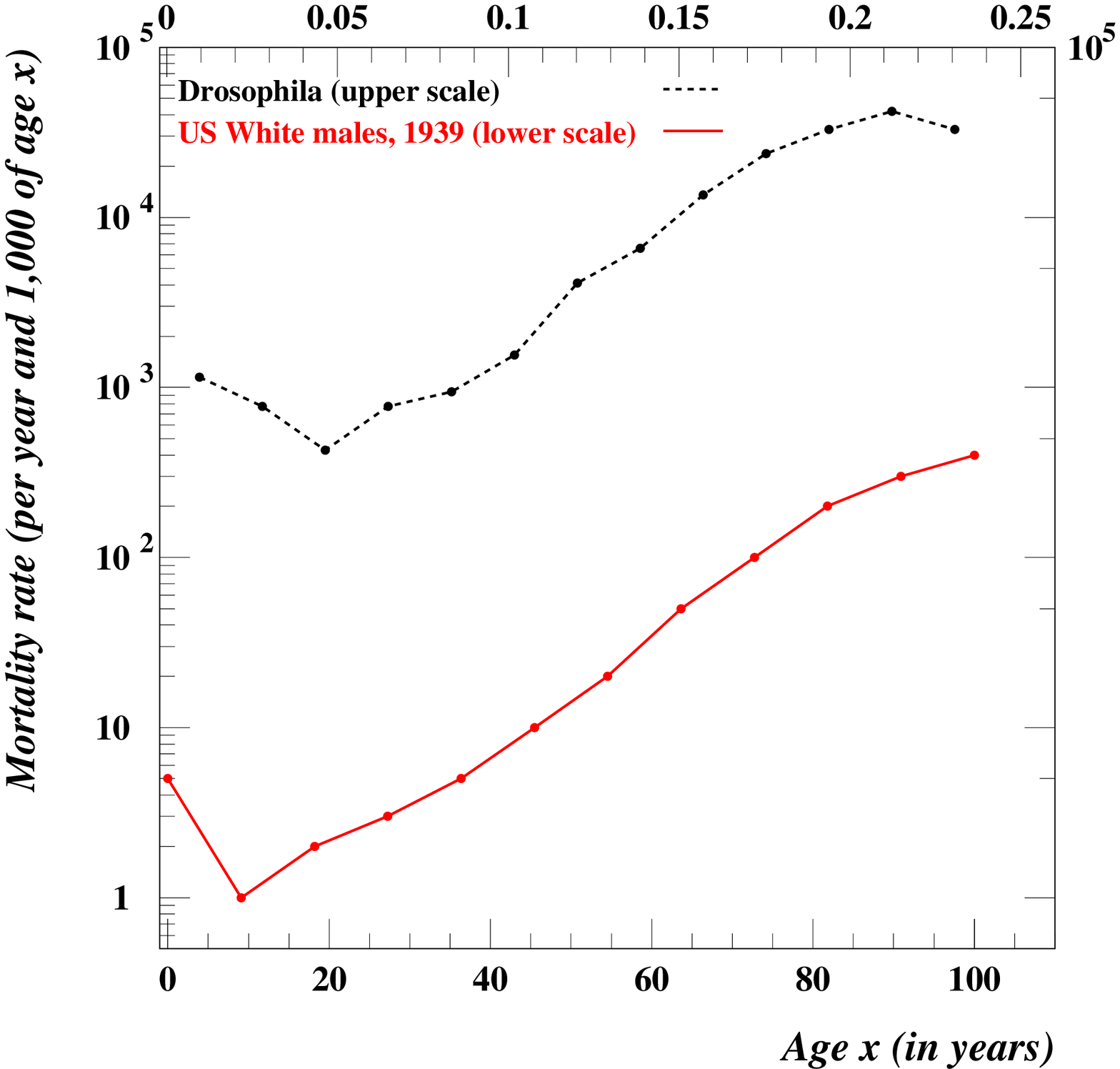}}
\qleg{Fig. \qhu 7\qhv Age-specific mortality rates for
humans and drosophila.}
{Gompertz's law holds in both cases.
As seen at the beginning of the
paper, the leveling off observed in humans should be
attributed to statistical fluctuations due to the
small numbers of the survivors. For the leveling off
of the drosophila curve we do not yet know whether it is
spurious or real.}
{Sources: Strehler (1967); Miyo and Charlesworth (2004),
Wang et al. (2013).}
\end{figure}

So far, the second point remains a conjecture. It was
tested through its consequences but direct verification will
become possible once reliable become available for populous
Asian countries. Needless to say, it is the birth dates which
pose a challenge because for present-day centenarians their birth
took place in a time where vital statistical records may not have
been completely accurate.
\qpar

Naturally, the analysis developed here for humans
can be extended to any other species for which
(i) Gompertz's law holds and (ii)
an upper bound of the life span can be demonstrated.
As an illustration let us consider the drosophila fruit flies.
As shown in Fig. 7,
their age-specific mortality rates
follow Gompertz's law.
\qpar
One may wonder whether
the leveling off seen in the drosophila curve 
is real or rather due to statistical fluctuations.
There is no doubt that the question could be settled
by performing repeated observations on
large populations. Anyway, 
it should be observed that there can
be an upper bound of the life span
even if there is a leveling off. Only a steady decrease of
$ \mu(t) $ would prevent it from reaching
the terminal rate of 1,000 per 1,000. 
\qpar
What can be expected from the observation of populations
of drosophila? The first question is of course to
see whether there is a fixed point or not
and if there is one how robust it is 
with respect to changes in
the parameters such as temperature, light, male/female ratio
which shape the living conditions of the drosophila.
It is obvious that if living conditions are not appropriate
the fixed-end point will remain out of reach. On the contrary,
conditions under which some
members of a cohort will reach the fixed-end point
will define the ``envelope'' of acceptable living 
parameters%
\qfoot{The word ``envelope'' is used in the same sense
as for the envelope of flight parameters of an aircraft.}%
.
Within this envelope it will be possible
to study the influence of various parameters, including
``social'' variables such as the male/female ratio
(in this respect see Carey et al. 2002 and Costa et al. 2010).
In this way, it may be possible to test the relationship
between $ d_1 $ and $ \alpha $ in an experimental way.
That should lead to clearer and more accurate results
than the observational method that one must use
for human populations.
\qpar

Naturally, the drosophila fruit flies
are not the only living organisms 
for which Gompertz's law holds. For instance, it was shown by
Raymond Pearl (1941) that the lifetables of
flour beetles are fairly similar to human life tables.

\qA{Comments on earlier papers}

The fact that the slope of Gompertz's law is country
and period dependent has been observed as soon as international
demographic data became available (e.g. Statistique 1907, INSEE 1954).
However, the explanations put
forward in such studies were very different from the one given
in the present paper. In a sense this is understandable because
the validity of Gompertz's law in the age range $ 95-106 $ 
which is a key-element in our explanation was clearly
established  only
in 1999 through the work of Thatcher and his collaborators.
\qpar
In the following lines we briefly discuss two studies:
Strehler (1960, 1967) and Beltr\'an-S\'anchez et al. (2012).
A third one, Finch (1990), is mentioned in the reference section.
\qpar
The international data used in Strehler (1960, 1967) are
from the UN Demographic Yearbook of 1955 and concern 32 countries.
Quite surprisingly the authors study the 
regression between $ \alpha $ and the ``logarithms of extrapolated
hypothetical mortality rates at age 0''. Is this not a case of
circular reasoning%
\qfoot{ The authors describe their procedure as follows: Each 
country's mortality rates [for various ages] were ``plotted on
semi-log paper. A straight line was drawn between points from 
35 to 85 (or 50 to 70 if large departures from linearity occurred),
and the slopes and intercepts were measured. Only a few countries,
whose Gompertz plots exhibited great irregularities were
excluded''. Thus, it seems clear that the intercepts 
were not drawn from an independent source.}%
?
In the present paper the death rates at age 37.5 were
taken from death data {\it independently} of the regression procedure. 
\qpar

The objective of the paper by Beltr\'an-S\'anchez et al. is
different. These authors wish to determine if there is a connection
between infant or child mortality, mid-age mortality and
the slope of Gompertz's law. This is indeed an important question.
The author show that between 1800 and 1915 there is indeed
a connection between early life mortality and late life mortality.
Why did they limit their study to 1800-1915? One can guess that
they left aside following decades because during
the 20th century medical progress brought about a dramatic
decline in early life mortality which would have made
the effect they wanted to study all too obvious. By limiting
themselves to the period prior to 1915, they have a better
chance to observe the biological aspect independently of
the impact of medical advances. Another way to achieve
that objective would have been to study this effect on
animal populations. This brings us back to the proposition
that if carried out in parallel with demographic studies,
the investigation of animal populations 
can provide valuable additional insight.

\vskip 10mm
{\large \bf References}
\qpar

{\color{blue} 
The objective of the comments which appear at the end of some 
of the entries is to indicate the implications of 
those studies for the present investigation. They may be removed
in the final version of the paper.}
\qpar

\qparr
Beltr\'an-S\'anchez (H.), Crimmins (E.M.), Finch (C.E.) 2012:
Early cohort mortality predicts the cohort rate of aging: 
an historical analysis.
Journal of Development Origins of Health and Disease 3,5,380-386.\qL
[This paper shows that mortality in the age interval $ (0,10) $
and mortality in the interval $ (40-90) $ (as described by
Gompertz's law) are connected.]

\qparr
Carey (J.R.), Liedo (P.), Harshman (L.), Zhang (Y.), M\"uller (H.-G.),
Partridge (L.), Wang (J.-L.) 2002: A mortality cost of virginity
at older ages in female Mediterranean fruit flies.
Experimental Gerontology 37,507-512.\qL
[One of the graphs shows that there is a high variability 
(coefficient of variation over 30\%)
in death rates in experiments on different cohorts even though
each cohort numbered some 1,000 flies. Altogether,
some 65,000 flies were used.
The paper suggests
that females reared without males have a higher death rate
than mating females except in the age interval $ 7-22 $ days.
The results are given without error bars.]

\qparr
Costa (M.), Mateus (R.P.), Moura (M.O.), Machado (L.P. de B.) 2010:
Adult sex ratio on male survivorship of {\it Drosophila
melanogaster}. Revista Brasileira de Entomologia 54,3,446-449.
[The purpose of this experiment was somewhat similar
to the objective of the Costa paper above. However,
only 42 males and 23 females were used and
it does not seem that the experiment was repeated several times. 
The results are given without error bars.]

\qparr
Finch (C.E.) 1990: Longevity, senescence and the
genome. Chicago University Press, Chicago.\qL
[This work shows that in 1990 the status of Gompertz's law
in old age was still unclear. On p. 15 one reads: 
``Deviations of slopes are often seen for survivors to very
advanced ages. While one could argue that the rate
of senescence decreases at later ages, another view 
is that the survivors to advanced ages are a special subpopulation
with special mortality statistics''.
This emphasizes the importance of the work of Thatcher et al. 
(1999) which solved this question by collecting an extensive
statistical database from many countries.]

\qparr
Institut National de la Statistique et des
Etudes economiques (INSEE) 1954: Le mouvement naturel de la
population dans le monde de 1906 \`a 1936. [International
vital statistics from 1906 to 1936]. Data compilation 
by Henri Bunle. \qL
[This volume contains data for mortality by age
including detailed neonatal and post-neonatal mortality data.]

\qparr
Institut National de la Statistique et des
Etudes economiques (INSEE) 1966: Annuaire Statistique de la
France {Statistical Yearbook of France]. \qL
[This yearbook is special in the sense that it includes
a retrospective summary of the statistics of previous decades.]

\qparr
Miyo (T.), Charlesworth (B.) 2004: Age-specific mortality rate
for Drosophila me\-lano\-gaster. Proceedings of Biological Sciences
of the the Royal Society, 7 December, 271,\-1556,\-2517-2522.

\qparr
Pearl (R.), Park (T.), Miner (J.R.) 1941: Experimental
studies on the duration of life. XVI. Life tables for the flour
beetle {\it Tribolium confusum Duval}.
The American Naturalist 75,756,5-19.\qL
[Only 800 beetles were used in this study. Yet,
the resulting life table curves seem much smoother (although no
error bars are given) than those shown
in Carey et al. (2002) in spite of the fact that 
as many as 66,000 fruit flies were used in that experiment.]

\qparr
Richmond (P.), Roehner (B.M.) 2015a: Effect of marital status
on death rates. Part I: High accuracy exploration of the
Farr-Bertillon effect. \qL
Submitted to Physica A. Available on the arXiv website at the
following address: \qL
http://lanl.arxiv.org/abs/1508.04939

\qparr
Richmond (P.), Roehner (B.M.) 2015b:  Effect of marital status
on death rates. Part II: Transient mortality spikes.\qL
Submitted to Physica A. Available on the arXiv website at the
following address: \qL
http://lanl.arxiv.org/abs/1508.04944

\qparr
Statistique G\'en\'erale 1907: 
Statistique internationale du mouvement de la population d'apr\'es les
registres de l'\'etat Civil.
R\'esum\'e r\'etrospectif depuis
l'origine des statistiques de l'\'etat civil jusqu'en 1909.
[Vital international statistics. Retrospective summary
from early vital statistical records to 1909.]
Vol. 1. National printing office.

\qparr
Strehler (B.L.), Mildvan (A.S.) 1960: General theory of
mortality and aging. Science 132,3418,14-21.\qL
[The subtitle is: ``A stochastic model relates observations
on aging, physiologic decline, mortality, and radiation''
which means that the aim of the theory is to explain
several facts rather than just one. Fig. 3 gives
a plot of the ``extrapolated mortality rate at age 0''
with respect to Gompertz's slope. The reason why
this is rather circular reasoning is explained in the text.]

\qparr
Strehler (B.L.) 1967: Mortality trends and projections.
Transactions of the Society of Actuaries 19,2,55,D429-D440.

\qparr
Thatcher (A.R.), Kannisto (V.), Vaupel (J.W.) 1999: 
The force of mortality at ages 80 to 120.
Odense University Press.\qL
[The authors have assembled a new ``Archive of Population Data on
  Aging''. It
  contains official statistics on deaths at ages 80
  and over in 30 countries.
The monograph is based on data for a sample of 13 countries,
mostly European countries and Japan:
Austria, Denmark, England and Wales, Finland,
France,  West Germany, Iceland, Italy, Japan, the   Netherlands,
Norway, Sweden, and Switzerland.
The United States is not included in the sample
because US statistics do not
give separate death rate data for age-groups older than 80-84.
Currently (i.e. in 2015) the population of these 13 countries
is about 450 millions.
In addition, the study pooled the data over a 30 year time interval,
namely 1960-1990. As a result, the database includes
120,000 persons who reached age 100.] 

\qparr
Wang (L.), Xu (Y.), Di (Z.), Roehner (B.M) 2013:
How does group interaction or its severance
affect life expectancy?
arXiv preprint (June 2013).

\end{document}